# ON THE "CRACKING" SCHEME IN THE PAPER "A DIRECTIONAL COUPLER ATTACK AGAINST THE KISH KEY DISTRIBUTION SYSTEM" BY GUNN, ALLISON AND ABBOTT

**Hsien-Pu Chen [1]), Laszlo B. Kish [1]), Claes G. Granqvist [2]), G. Schmera [3])**

[1]) *Texas A&M University, Department of Electrical and Computer Engineering, College Station, TX 77843-3128, USA*

[2]) *Department of Engineering Sciences, The Ångström Laboratory, Uppsala University, P.O. Box 534, SE-75121 Uppsala, Sweden*

[3]) *Space and Naval Warfare Systems Center, San Diego, CA 92152, USA*

**Abstract**

Recently, Gunn, Allison and Abbott (GAA) [http://arxiv.org/pdf/1402.2709v2.pdf] proposed a new scheme to utilize electromagnetic waves for eavesdropping on the Kirchhoff-law–Johnson-noise (KLJN) secure key distribution. We proved in a former paper [Fluct. Noise Lett. 13 (2014) 1450016] that GAA's mathematical model is unphysical. Here we analyze GAA's cracking scheme and show that, in the case of a loss-free cable, it provides less eavesdropping information than in the earlier (Bergou)-Scheuer-Yariv mean-square-based attack [Kish LB, Scheuer J, Phys. Lett. A 374:2140–2142 (2010)], while it offers no information in the case of a lossy cable. We also investigate GAA's claim to be experimentally capable of distinguishing—using statistics over a few correlation times only—the distributions of two Gaussian noises with a relative variance difference of less than $10^{-8}$. Normally such distinctions would require hundreds of millions of correlations times to be observable. We identify several potential experimental artifacts as results of poor KLJN design, which can lead to GAA's assertions: deterministic currents due to spurious harmonic components caused by ground loops, DC offset, aliasing, non-Gaussian features including non-linearities and other non-idealities in generators, and the time-derivative nature of GAA's scheme which tends to enhance all of these artifacts.

## 1. Introduction

Recently Gunn, Allison and Abbott (GAA) [1] proposed a new scheme to utilize electromagnetic waves for eavesdropping on the Kirchhoff-law–Johnson-noise (KLJN) secure key distribution. In a former paper [2], we proved that claims concerning electromagnetic waves are unphysical in GAA's attack, since the quasi-static limit holds for the KLJN scheme and implies that physical waves do not exist in its wire channel. An assumption of wave modes in a short cable, and in the low-frequency limit, in fact violates a number of laws of physics, including the Second Law of Thermodynamics. One aspect related to these mistakes is that, in electrical engineer jargon, all oscillating and propagating time functions are called waves, whereas in physics the corresponding retarded potentials can be wave-type or non-wave-type. Physical waves involve two dual energy forms that regenerate each other during propagation; these forms can involve electrical and magnetic fields, or deal with kinetic and potential energy as in the case of elastic waves. Non-wave-type retarded potentials in the quasi-static regime, however, have negligible crosstalk between the two energy forms, and energy exchange takes place between them and generators [2]; this latter situation pertains to the KLJN scheme. We note in passing that, while there are no physical waves in the KLJN system, the propagation delay of the non-wave-type retarded potentials may still provide information for Eve, and therefore a correct analysis is essential.

In the steady-state driving case, the correct analysis [2] shows that the starting d'Alembert equation

$$U(t,x) = U_+\left(t - \frac{x}{v}\right) + U_-\left(t + \frac{x}{v}\right),$$ (1)

which is the foundation of GAA's approach, is invalid because the system under study cannot be described with a single phase velocity [2], but these velocities are directionally dependent during secure key exchange. Here $U_+$ and $U_-$ are voltage components of waves propagating to the right and left along the $x$-axis, and $v$ is a single propagation velocity. GAA used Eq. 1 to deduce the equations

$$\frac{dU}{dt} + v\frac{dU}{dx} = 2\frac{dU_+}{dt}$$ (2)

and

$$\frac{dU}{dt} - v\frac{dU}{dx} = 2\frac{dU_-}{dt}$$ (3)

as a basis of their "directional coupler" attack. Their claim [1] is that the quantities at the left-hand side of Eqs. 2 and 3 are measurable so that the time derivatives at the right-hand side of the equations can be calculated and used for eavesdropping.

Before analyzing the experimental claims and potential artifacts, we take a closer look at the mathematics of Eqs. 1–3.

## 2. Mathematical analysis of GAA's scheme

In this section, we present a correct analysis of GAA's scheme and show that Eve's eavesdropped information is always less within the GAA scheme than within the old mean-square attack based on the comparison of two end-voltages [3], unless there are flaws in the realization of the KLJN key exchanger.

We assume in the rest of the paper that the bit-value arrangement between Alice and Bob is mixed, *i.e.*, one of them connects the large resistance to the cable and the other uses the small resistance. This situation indicates not only a secure key exchange event but also that different phase velocities must be used for the two directions in Eq. 1 during steady-state conditions (see also related theory and verifications by simulation in our former paper [2]).

### 2.1 General considerations

Even for waves, Eq. 1 is not suitable for steady-state excitations [4] and the second term violates causality. However there is a way to modify this equation under steady-state conditions in the case of KLJN by using direction-dependent phase velocities [2] of retarded potentials. Furthermore causality is ascertained by setting

$$U(t,x) = U_+\left(t - \frac{x}{v_+}\right) + U_-\left(t - \frac{D-x}{v_-}\right),$$ (4)

where $x = 0$ and $x = D$ pertain to the left-hand and right-hand ends of the cable, respectively, and $D$ is cable length. The phase velocities are

$$v_+ = \frac{DR_B}{L_c} \quad \text{and} \quad v_- = \frac{DR_A}{L_c} \quad , \tag{5}$$

where $R_A$ and $R_B$ are Alice's and Bob's resistances, $L_c$ is the inductance of the cable and it is assumed that Alice is at $x = 0$ and Bob is at $x = D$.

It is important to realize that, according to Eqs. 4 and 5, Eve must know Alice's and Bob's resistor values in order to have the correct input for the GAA experiments. This implies that Eve's one-bit uncertainty persists, which is the *indication* of security. A *proof* of security will be given below in Eq. 21.

Our earlier work [2] proved that, in the quasi-static frequency limit pertinent to the KLJN scheme, the exact distributed-impedance rendition of the cable shown in Fig. 1 leads to the simplified serial impedance models in Figs. 2a and 2b because the capacitive currents converge towards zero in the limit of low frequencies. Figure 2a is a first-order approximation of the real situation while Fig. 2b models a situation wherein the cable is lossless or the voltage drop on the resistive component is negligible compared to that of the inductive component in the dominant frequency range of the quasi-static regime.

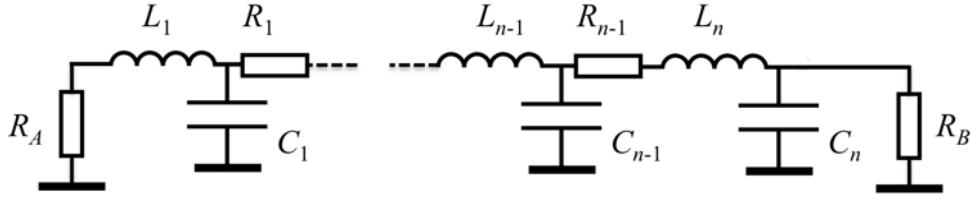

**Figure 1.** Outline of the pertinent part of the KLJN scheme with a distributed LCR model of a long and leakage-free cable [2]. When the cable losses can be neglected, one may omit the $R_i$ resistors representing the distributed resistance of the cable. Alice's and Bob's resistors—denoted $R_A$ and $R_B$, respectively—are randomly selected from the set $\{R_L, R_H\}$ with $(R_L \neq R_H)$ at the beginning of each bit-exchange period. These resistors, with associated serial generators (not shown), emulate thermal noise with high noise temperature and strongly limited bandwidth.

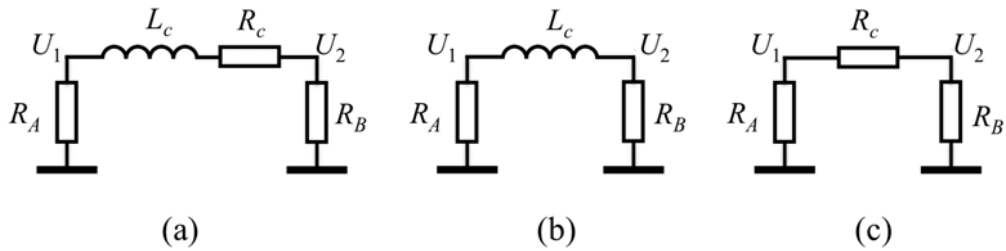

**Figure 2.** Lumped impedance-components-based model of a short cable at low frequencies for analyzing voltage drop along the cable and phase shift in the quasi static limit [2]. Part (a) represents a cable with loss (cable inductance and resistance are designated $L_c$ and $R_c$, respectively), and part (b) represents a lossless cable. Part (c) is used to determine the voltage drop in the asymptotic case where loss dominates the cable impedance (this case is not practical and is used only for the sake of analysis).

Figure 3 provides a starting point for our security analysis of the KLJN scheme. $U_A$ and $U_B$ are voltages of the (thermal) noise voltage generators, and $R_A$ and $R_B$ are Alice's and Bob's resistors, respectively. Here $U_1$ and $U_2$ are voltages at the two ends of the cable, and $Z_c$ denotes cable impedance. Furthermore we set $U_{12} = U_1 - U_2$.

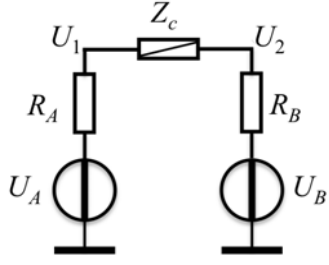

**Figure 3.** Circuit for impedance-based analysis of GAA's attack, as described in the main text.

In order to accomplish security, the voltage drop on the cable must be kept small in comparison with other voltages [3], *i.e.*,

$$U_1 \cong U_2 \equiv U \quad . \tag{6}$$

We first consider the earlier mentioned wire-resistance-based attack [3,5], wherein miniscule differences between the mean-square voltages $\langle U_1^2 \rangle$ and $\langle U_2^2 \rangle$ served as information leak toward Eve. In the experimental demonstration, [5], chossing the number of observed correlation times $N_{oc}$ during bit-exchange to 50, the wire resistance to 200 Ω, $R_A = 2$ kΩ and $R_B = 9$kΩ, resulted in Eve's successful guessing probability $p$=0.525, which means that the relative information leak was 0.19% of the exchanged key bits. This implies that two-stage privacy amplification would be needed [6] to reduce this leak to a level below the desired one of $10^{-8}$. For this type of attack, $p$ scales as [7]

$$p = 0.5 + \theta \frac{|Z_c|^2}{R_A R_B} \tag{7}$$

at fixed $N_{oc}$, where $\theta$ is a constant which depends only on $N_{oc}$.

Turning now to GAA's experiments [1], $|Z_c|^2$ is about $10^5$ times smaller than before while their $R_A$ and $R_B$ have similar values (1 kΩ and 10 kΩ, respectively). With the same $N_{oc}$ as above and using the old method [3] together with GAA's parameters, Eve's probability of successful bit guessing would be [6]

$$p \approx 0.5000002 \quad , \tag{8}$$

which is better than even the value of $p$ needed to secure the upper limit of $10^{-8}$ for the relative information leak ($p = 0.5006$ [6]).

In stark contradiction to the results above, GAA assert that by using their standard statistical method they measure

$$p \approx 1 \tag{9}$$

at the given conditions. This is an extraordinarily strong claim which—if correct—would mean that Eve can perform a nearly-deterministic guess not only about the bit-states but also of the exact time dependence of Alice's and Bob's noise voltages.

### 2.2 Case 1: Lossless short cable with very small impedance

In the case of a *lossless cable* with very small impedance, we assume that Rel. 6 holds but that $U_{12}$ remains measurable. Suppose now that Eve employs Eq. 2 to extract information. Using the proper velocity and the measurable quantities at the left-hand side of Eq. 2a, we get

$$\frac{dU}{dt} + v_+ \frac{dU}{dx} = 2 \frac{dU_x}{dt} \quad . \tag{10}$$

The resulting voltage $U_x(t)$ in the right-hand side of Eq. 2a needs clarification at this point. To this end we Fourier transform Eq. 6 and obtain

$$j\omega U(\omega) + v_+ \frac{dU(\omega)}{dx} = 2j\omega U_x(\omega) \quad , \tag{11}$$

so that

$$2U_x(\omega) = U(\omega) + \frac{v_+}{j\omega} \frac{dU(\omega)}{dx} \tag{12}$$

or

$$2U_x(\omega) = U(\omega) + \frac{DR_B}{j\omega L_c} \frac{dU(\omega)}{dx} = U(\omega) + \frac{DR_B}{j\omega L_c} \frac{U_{12}(\omega)}{D} \quad , \tag{13}$$

where $U_{12}(\omega) = U_1(\omega) - U_2(\omega)$ . Using Ohm's law for impedance, we have

$$2U_x(\omega) = U(\omega) + R_B \frac{U_{1,2}(\omega)}{j\omega L_c} = U(\omega) + R_B I(\omega) \quad . \tag{14}$$

A relation

$$U(\omega) \cong U_1(\omega) \cong U_2(\omega) \quad , \tag{15}$$

which is the Fourier transform of Eq. 6, holds when the cable impedance is very small (*cf.* Fig. 3) so that

$$2U_x(\omega) = U_2(\omega) + R_B I(\omega) = U_B(\omega) \quad . \tag{16}$$

After inverse Fourier transformation and substituting the voltages back into Eq. 6, it is found that the corrected Eq. 2a reads as

$$\frac{dU}{dt} + v_+ \frac{dU}{dx} = \frac{dU_B(t)}{dt} \quad . \tag{17}$$

Similar considerations for GAA's other equation, with the opposite sign of the second term, lead to

$$\frac{dU}{dt} - v_- \frac{dU}{dx} = \frac{dU_A(t)}{dt} \quad . \tag{18}$$

The right-hand sides of Eqs. 17 and 18 give the time derivative of the voltages of Alice's and Bob's generators provided Eve uses the *correct guess* and consequently substitutes the *correct resistances* in these equations. This result proves that GAA do not have a directional coupler but something else, which can be called a "separator" and is able to extract the voltage amplitudes of Alice's and Bob's generators (without the voltage-division caused by the resistor at the other end). Such a tool would be even better for Eve, but it works only if the correct phase velocity is assumed. The phase velocity in the steady state is determined by the unknown resistor terminating the cable toward the propagation direction [2], and therefore

Eve must correctly guess the value of the resistor at that end in order to obtain the correct voltage.

The most important question is this: what happens if Eve assumes the *wrong resistor value* at Bob's side, *i.e.*, if Eve assumes Alice's resistor value? Obviously, the resulting voltage $U_x(t)$ will then be a weighted superposition of the voltages seen by Alice and Bob. However, the real question concerns the statistical properties associated with Eve's choice. Can Eve utilize these properties to extract information?

The answer to the questions is simple, and we first observe that the voltage $U_x(t)$ will be a Gaussian noise [8–10], because a linear combination of Gaussians results in a Gaussian as a consequence of the Central Limit Theorem [11]. Thus the real question regards the variance of the voltage. Its calculation is straightforward, and Eq. 16 becomes

$$2U_x(\omega) = U_2(\omega) + R_A I(\omega) \cong U_1(\omega) + R_A I(\omega) \quad . \tag{19}$$

Here it is important to realize that the cable voltage and cable current are orthogonal—*i.e.*, uncorrelated—in order to ensure zero net power flow and satisfy the Second Law of Thermodynamics [7,12–15], so that

$$\langle U(t)I(t) \rangle = 0 \quad . \tag{20}$$

Thus Pythagoras' Rule gives that the variance (mean-square) of the sum at the right-hand side of Eq. 19 is invariant to changing the plus sign to a minus sign, which is given a pictorial rendition in Fig. 4. It follows that

$$\langle U_x^2(t) \rangle = \langle U_1^2(t) \rangle + R_A^2 \langle I^2(t) \rangle = \langle U_1^2(t) \rangle - R_A^2 \langle I^2(t) \rangle \quad , \tag{21}$$

which is exactly the variance of Alice's noise voltage in accordance with Kirchhoff's law (*cf.* Fig. 3). It should be remarked that GAA used time derivatives, but this does not change the situation of orthogonality.

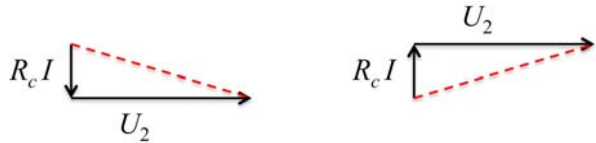

**Figure 4.** Illustration showing that added orthogonal noise voltages produce the same RMS voltage and mean-square voltage even if the sign of the current is flipped to the opposite value. The resulting time-dependent voltages will be different although their RMS and mean-square and RMS amplitudes remain the same.

The mean square voltage always corresponds to that of the noise source of the assumed resistor, which means that Eve gets what she assumes instead of learning about the true bit-situation provided Rel. 6 holds. Thus the only role of the inductance of the lossless cable is to detect the current in the wire, so that Eve's one-bit inaccuracy remains which proves the security of the key exchange against this attack. It should also be noted that "separators" of the above mentioned kind can easily be realized by directly measuring the current and using Ohm's law with guessed resistance values to determine the voltages at Alice's and Bob's ends. A further discussion of the latter issue was given elsewhere [7], where the separators were described and referred to as "impedance-based directional couplers", and where it was pointed out that they are useless for Eve. Thus the obtained mean-square voltages satisfy the *supposed* resistance value, and Eve cannot extract any information by using this system provided Rel. 6 holds. Section 2.4 below elucidates the role of the approximation leading to Rel. 6.

### 2.3 Case 2: Short cable dominated by loss

For a *lossy cable*, the voltage drop over the resistor makes even the modified d'Alembert-equation-approach in Eq. 4 invalid, even if the correct phase velocity is used. Equations 12 and 13 become

$$2U_x(\omega) = U(\omega) + \frac{v_+}{j\omega}\frac{U_{12}(\omega)}{D} = U(\omega) + \frac{DR_B}{j\omega L_c}\frac{U_{12}(\omega)}{D} = U(\omega) + \frac{R_B}{L_c}\frac{U_{12}(\omega)}{j\omega} \quad , \tag{22}$$

and inverse Fourier transformation yields

$$2U_x(t) = U(t) + \frac{R_B}{L_c}\int U_{12}(t)dt \quad . \tag{23}$$

The obtained $U_x$ does not have any meaning or information for Eve, because $U(t)$ and the integral of $U_{12}(t)$—which is proportional to the time integral of the current—are orthogonal even if the current and $U(t)$ have some small correlation due to loss.

### 2.4 Conclusion of sections 2.2 and 2.3

As shown above Eve cannot extract any information, neither in the lossless nor in the lossy cable, provided Rel. 6 holds. However, this relation is only approximate, because there is a non-zero difference $U_{12}$ between $U_1$ and $U_2$. This small difference causes a small offset between the related results which indeed is information for Eve. But this offset is the very same as that utilized directly in the old mean-square-comparison based wire-resistance-attack method [3] without the extra noise components shown above. The extra noise components weaken Eve's information, and therefore the conclusion is straightforward: GAA's method always provides less information than the old wire-resistance-attack [3].

## 3. Experiments: What could go wrong?

Many things could go wrong in GAA's experiments claimed to prove the validity of their attack against the KLJN scheme. Here we try to identify the most probable deficiencies but presume that conceptual errors concerning the experiments are not present. From the many possibilities, we select only a few and only those directly related to the realization of the KLJN scheme but not to the measurement set-up as such.

### 3.1 The experimental claim

GAA [1] used a standard statistical method to compare distributions of extracted voltage components and to identify the bit (resistor) arrangement at the two ends of the wire in the KLJN scheme. They asserted that they were able to identify the resistor arrangement within a very short time in the case of lossy cables.

Let us now estimate the observable relative difference of the mean-square voltages at the two cable ends in GAA's experiment: The resistors were 1 kΩ and 10 kΩ, and the cable length was 1.5 m and 2 m. GAA did not specify their cable parameters, but a 1.5-meter-long cable taken to be, as a reasonable assumption, a copper wire with a cross section of 1 mm$^2$ yields a cable resistance of 0.07 $\Omega$.

As seen above, the old wire-resistance-attack [3] gives an upper limit for the extracted information. A mean-square operation is an efficient estimator for Gaussian processes [8], and therefore other statistical methods cannot offer much advantage. Using the result in an earlier paper [3] for the measurable relative mean-square voltage difference, we obtain

$$\Delta_{rel}^2 \approx \frac{\left| \left\langle U_1^2(t) \right\rangle - \left\langle U_2^2(t) \right\rangle \right|}{\left\langle U_1^2(t) \right\rangle} \approx \frac{\left| \left\langle U_1^2(t) \right\rangle - \left\langle U_2^2(t) \right\rangle \right|}{\left\langle U_2^2(t) \right\rangle} \approx \frac{R_c^2}{R_A R_B} = \frac{0.07}{10^3 10^4} = 7 \times 10^{-9} \quad , \qquad (24)$$

showing that the imbalance of the mean-square voltages of the two Gaussian noises is less than $10^{-8}$. GAA's claim to identify which one of these distributions is the narrower by sampling a few correlation times is untenable, and normally hundreds of millions of correlation times would be required for a reasonably low error probability.

The question then arises as to what GAA did measure and how they obtained their surprising results?

### 3.2 Non-Gaussianity

According to the security proofs in earlier work [9,10], it is a strict mathematical requirement for the security of the KLJN scheme to have Gaussian processes, which means that the time derivatives also must be Gaussians. GAA did not specify their waveform generator, and thus the degree of Gaussianity remains unclear.

It is important to notice that most commercial noise generators use algorithms and filtering to approach Gaussians. Due to the Central Limit Theorem [11], time-integration shifts the statistics of noises toward Gaussians whereas time derivatives, which were used by GAA, strongly amplify non-Gaussian components.

Thus one strong candidate for causing the poor performance of the KLJN system in GAA's study [1] is non-Gaussianity of time derivatives.

### 3.3 Aliasing effects, non-linearity, and spurious noise components

Aliasing effects (which cause high-frequency non-Gaussian noises), non-linearity and other types of spurious signals in the generator are other strong candidates for destroying Gaussianity. Again, a time derivative will severely emphasize these weaknesses.

### 3.4 Deterministic currents in the loop

A low-frequency or DC current component may exist in the cable of the KLJN scheme and may be caused by a ground loop (leading to a 50/60Hz sinusoidal current) or DC offset. The voltage drop originating from such parasitic currents will introduce a location-dependent bias into the key distributions and quickly uncover the nature of the resistors at the two ends of the wire, as illustrated in Fig. 5.

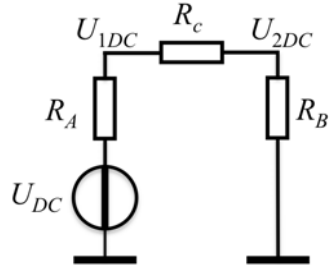

**Figure 5.** Schematic circuit for illustrating parasitic DC and low-frequency artifacts caused by ground loops. For the sake of simplicity, the parasitic source is assumed to exist solely at Alice's side. Only the parasitic voltage generator is shown, because its impact is additive to the voltages caused by other circuit deficiencies. The parasitic DC or low-frequency components $U_{1DC}$ and $U_{2DC}$ of $U_1$ and $U_2$, respectively, are sensitive to the location of the low/high resistor at Alice's and Bob's side.

However, Eve does not need to use GAA's method [1] to elucidate the resistor values: She can simply measure and compare the DC or 50/60 Hz voltage components of the strongly correlated voltage noises at the two ends of the wire and extract the key or its inverse. Figure 6 shows, as an example, computer simulations of two strongly correlated noises with a small DC shift. In this particular case, a single-time measurement is able to identify the DC voltage shift and uncover the key or its inverse. If the DC shift is greater than the stochastic difference between the time functions, then a single-time measurement is sufficient to distinguish the two noises and the bit-situations in the KLJN scheme. Concerning a 50/60 Hz parasitic signal in the loop, the period duration is about a hundred times greater than the correlation time of the noise with 5 kHz bandwidth used by GAA and thus, during the few correlation times used by GAA, this disturbance behaves practically as a parasitic DC shift.

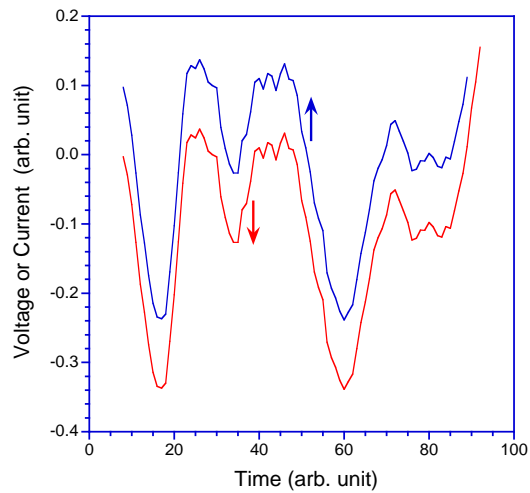

**Figure 6.** Computer-generated illustration of how a DC shift can distinguish between two strongly correlated noises by comparison at a single moment of time. The arrows indicate the directions of shift.

For the situation illustrated above, GAA's result, that Eve's successful guessing probability is progressively enhanced by increasing cable loss, is obvious. A lossless cable is represented by an inductance (*cf*. Fig. 2b) which produces a voltage drop proportional to the time derivative of the current. This means zero DC voltage and zero shift between the distribution functions due to a parasitic DC current in the loop. However the situation is changed when cable loss is present, and then a DC voltage shift will occur in accordance with Ohm's law as a result of the cable resistance $R_c$. This effect will be strongly enhanced by the time derivation of the channel voltage in GAA's scheme because the voltage drop in the cable is not time-derivated.

## 4. Conclusion

We have shown that GAA's approach [1] is invalid and that their experimental results must be caused by artifacts. Nevertheless, a correct interpretation of GAA's results is very enlightening because it shows clearly that nonlinearities, non-Gaussianity (even a weak one), aliasing effects and parasitic currents constitute very dangerous potential non-idealities in a practical KLJN system. The removal of such effects is straightforward, however, and can be accomplished by careful circuit design, filters, etc., while ignoring these effects can lead to cracking of the key. Furthermore, a well-defended KLJN system can execute spectral and statistical analysis on the noise in the cable and, together with a proper computer model of the cable, ascertain that effects due to parasitic currents are not present and thereby assure safe results. These types of checks are possible because the KLJN system is a classical physical one, and classical physics permits continuous monitoring of signals and parameters of the channel without destroying this information, which is a situation very unlike that in a quantum system. Thus the robustness of the KLJN state is offered by classical physics and is essential for the security of the key exchange within the KLJN scheme.

## Acknowledgements

Discussions with Janusz Smulko and Robert Nevels are appreciated.